\documentclass[twocolumn]{elsart5p}

\usepackage{natbib}

\usepackage{graphicx}
\usepackage{amssymb}
\usepackage{amsmath}

\usepackage{lineno}

\begin{document}

\begin{frontmatter}



\title{Metadynamic sampling of the free energy landscapes of proteins coupled with a Monte Carlo algorithm}


\author[uni]{F. Marini\corauthref{cor}},
\corauth[cor]{Corresponding author.}
\ead{franz.marini@mi.infn.it}
\author[uni,infn]{C. Camilloni},
\author[uni,infn]{D. Provasi},
\author[uni,infn,niels]{R.A. Broglia},
\author[uni,infn]{G. Tiana}
\address[uni]{Department of Physics, University of Milan, via Celoria 16, 20133 Milan, Italy}
\address[infn]{INFN, via Celoria 16, 20133 Milan, Italy} 
\address[niels]{The Niels Bohr Institute, University of Copenhagen, Blegdamsvej 17, DK 2100 Copenhagen, Denmark}

\begin{abstract}
Metadynamics is a powerful computational tool to obtain the free energy landscape of complex systems. The Monte Carlo algorithm has proven useful to calculate thermodynamic quantities associated with simplified models of proteins, and thus to gain an ever--increasing understanding on the general principles underlying the mechanism of protein folding. We show that it is possible to couple metadynamics and Monte Carlo algorithms to obtain the free energy of model proteins in a way which is computationally very economical.
\end{abstract}


\end{frontmatter}

% main text
\section{Introduction}
\label{intro}

Metadynamics is an algorithm which coupled to molecular dynamics provides an efficient tool to obtain the energy landscape of systems displaying large energy barriers, and thus whose sampling by standard tools is, at best, problematic. It is based on the knowledge of few slow collective variables $s_i$ of the system and on the use of a non--Markovian potential $U(s_i)$ that disfavors the exploration of regions of the phase space already visited by the system (\cite{Lai.Par:02}). This algorithm has been succesfully used to obtain the free energy of molecular systems at atomic detail (\cite{Bab.ea:06}).

In the case of simplified protein models, where the atomic structure of each amino acid is coarse--grained, it is common to sample the conformational space with the help of Monte Carlo algorithms. Such an approach is computationally more economic and more simple to implement than the corresponding molecular dynamics algorithm (see, e.g. \cite{Shi.ea:01}, \cite{Kus.ea:02}, \cite{Shi.Sha:02}). It is then natural to try to extend metadynamics so as to make it possible to couple it to a Monte Carlo algorithm.

Of course, other modifications of the straight Monte Carlo sampling have been developed during the last tens of years, including simulated tempering, multicanonical sampling, parallel tempering, etc. All of them are aimed at preventing the system to get trapped in free energy minima. In the following we show that Monte Carlo metadynamics is efficient, accurate and particularly easy to implement.

\begin{figure}[htb!]
\begin{center}
\includegraphics*[width=0.45\textwidth]{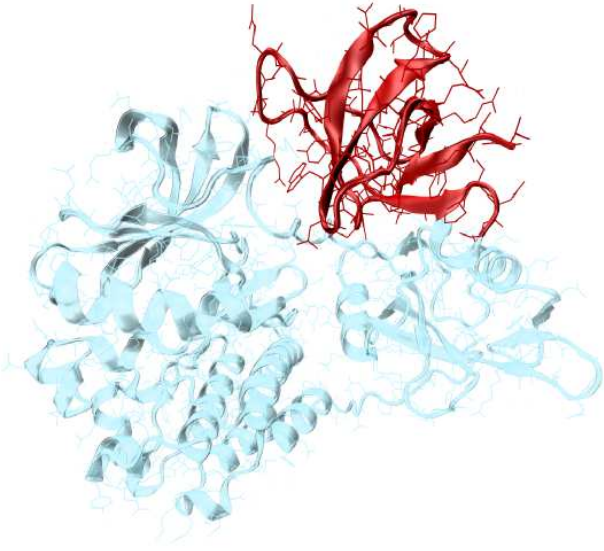}
\caption{Src protein (\emph{Proto-oncogene tyrosine-protein kinase Src}). The upper right part (dark, red online) is the SH3 domain.}
\label{fig:allsrc}
\end{center}
\end{figure}

We apply a scheme to the calculation of the free energy, as a function of the RMSD, of a small domain protein, namely Src-SH3. It is a widely studied domain (\cite{Gra.ea:98},\cite{Yi.ea:98},\cite{Rid.ea:99}) of the \emph{Proto-oncogene tyrosine-protein kinase Src}, a 536 residue protein that plays a multitude of roles in cell signalling.  Src is involved in the control of many functions, including cell adhesion, growth, movement and differentiation. SH3 is a domain built out of 60 residues, displaying mainly $\beta$-strands (see Fig. \ref{fig:allsrc}).
From calorimetry and fluorescence experiments, it is known to fold according to a two--state mechanism, that is, populating at biological temperature mainly two states (the native and the unfolded state) (\cite{Gra.Bak:97}). Consequently, we expect the free energy landscape to display two minima separated by a barrier. 

In Section \ref{method} we present the protein model used in the simulations along with a working description of the algorithm. We devise a formal proof of the correctness of the method in Section \ref{theory} and then test it in the specific case of Src-SH3 in Section \ref{results}.

\section{Method}
\label{method}

\begin{figure}[htb!]
\begin{center}
\includegraphics*[width=0.45\textwidth]{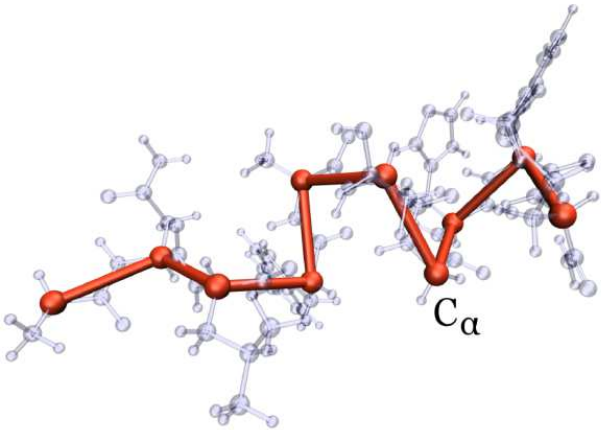}
\caption{Schematic protein model used in the simulations. The whole protein is shown, and in particular the linear C$_\alpha$ chain (dark, red online). Also visible is the sidechain (light blue online), which is although not used in the model.}
\label{fig:model}
\end{center}
\end{figure}

The model employed in the simulations describes the protein as a chain of beads centered on the C$_\alpha$ of the protein backbone (see Fig. \ref{fig:model}).
The allowed moves are the \emph{flip--move} and \emph{tail--flip}. The interactions are described by a G{\={o}}--model (\cite{Go:75}), where the only contacts participating in the potential energy calculation are the native contacts.

Every $\tau$ steps of the Monte Carlo sampling (\cite{Met.Ula:49}), the non-Markovian energy contribution is updated by adding a Gaussian hill with height $W$ and spread $\delta s$, centered around the current values of the collective variables.

Each Monte Carlo step, we apply a \emph{Metropolis} algorithm (\cite{Met.ea:53}) where the transition probability is given by
\begin{equation}
\begin{split}
w({\mathbf x}_n \rightarrow {\mathbf x}_{n^\prime}) &= w_0 p_{ap}(n \rightarrow n^\prime) \\
         &\times {\rm min}\left[ 1, {\rm e}^{-\frac{E^\prime - E + U(s_i)^\prime - U(s_i)}{k_B T}} \right] \;, \\
\end{split}
\label{eqn:metropolis}
\end{equation}
that is, the probability with which the next Monte Carlo move is accepted is calculated on the variation of the energy of the system, plus the variation of the metadynamics potential.

\section{Theory}
\label{theory}

During each fragment of trajectory after the update of the non--Markovian potential at each time $T$, the collective variable $s$ explores a region $A(T)$. If one makes the critical assumption that the dynamics has been able to visit this region so extensively that ergodicity holds, then the probability distribution of the collective variable is
\begin{equation}
P(s,T)=\frac{\exp[-\beta(F(s)+U(s,T))]}{\int_{A(T)}ds'\;\exp[-\beta(F(s')+U(s',T))]} \;.
\label{eq_pst}
\end{equation}
After the end of this sampling, the non--Markovian potential is updated, and the new potential reads
\begin{equation}
U(s,T+\tau) = U(s,T) + W \tau P(s,T)\; ,
\label{eq_ust}
\end{equation}
where $W$ is the heigth of the energy added to the non--Markovian potential. Further assuming that $W$ is small, that is that the new term does not perturb in an important way the shape of the potential $U$, then the previous equation can be rewritten as
\begin{equation} 
\begin{split}
\frac{dU(s,T)}{dT} &= \\
                   &W \frac{\exp[-\beta(F(s)+U(s,T))]}{\int_{A(T)}ds'\;\exp[-\beta(F(s')+U(s',T))]} \;. \\ 
\end{split}
\label{eq_du}
\end{equation}
Once the free energy landscape is completely filled by the non--Markovian energy, then the growth of this non--Markovian energy will be independent on $s$, that is $dU/dT=W/A$, or equivalently
\begin{equation}
\begin{split}
\exp[-\beta(F(s) &+ U(s,\infty))] =\\
                 &\frac{1}{A}\int_{A}ds'\;\exp[-\beta(F(s')+U(s',\infty))] \;, \\
\end{split}
\end{equation}
where $A$ is the whole interval spanned by the collective variable. Integrating by saddle--point evaluation, leads to
\begin{equation}
\begin{split}
F(s) &= \\
     &- U(s,\infty) -\beta^{-1}\log\left(\frac{1}{A}\sqrt{\frac{2\pi}{\beta|F''(s_0)+U''(s_0)|}}\right) \\
     &+ F(s_0)+U(s_0) \;, \\
\end{split}
\end{equation}
where $s_0$ is defined by $U'(s_0)=-F'(s_0)$. The last equation states that the free energy of the system is, except for an additive constant, equal to the opposite of the non--Markovian potential. A nice property of this algorithm is that the obtained free energy depends logarithmically on any additive error in the determination of $U(s,T+\tau)$ (i.e., if one adds $\epsilon(s) \tau$ to Eq. (\ref{eq_du}), one obtains an additive term $\log\epsilon(s)$ in $F(s)$).

\section{Results}
\label{results}

\begin{figure}[htb!]
\begin{center}
\includegraphics*[width=0.45\textwidth]{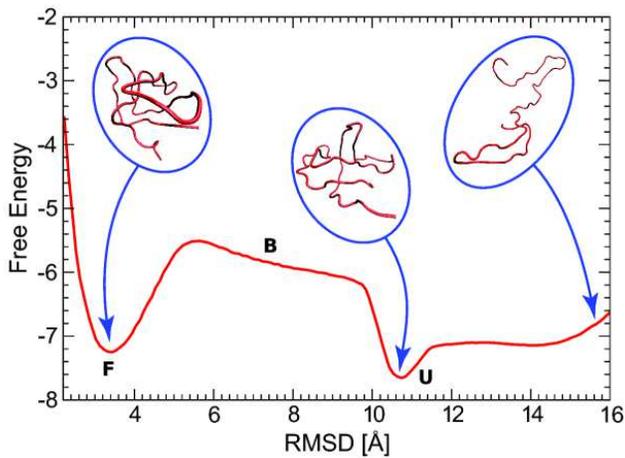}
\caption{Reference energy landscape as a function of the RMSD, obtained from a 90 billions of step long Monte Carlo simulation. Shown in the insets are a typical folded configuration (\textbf{F}), a compact unfolded one (\textbf{U}) and an elongated unfolded one (right).}
\label{fig:reference_landscape}
\end{center}
\end{figure}

In order to obtain a reference free energy landscape as a function of the RMSD for comparison to the metadynamics reconstructed landscapes, we first carried out a fairly long classical Monte Carlo simulation (90 billions of Monte Carlo Steps (MCS)). The free energy calculated at temperature $\theta=0.625$ (slightly below the folding temperature, defined as the temperature at which the volume of the native basin is equal to that of the denatured basin) is displayed in Fig. \ref{fig:reference_landscape}. After 80 billion steps the root mean square difference between the landscape at time $T$ and at time $T-\Delta T$, where $\Delta T=10000$ MCS, was constantly below $10^{-2}$ \AA, indicating that the free energy is likely to have reached its equilibrium shape. 

The landscape presents a fairly broad barrier between the folded and the unfolded states (marked with \textbf{B} in Fig. \ref{fig:reference_landscape}). Also shown in Fig. \ref{fig:reference_landscape} are a typical folded configuration (the shown configuration has a RMSD of 3.11 \AA, less than the distance between two consecutives C$_\alpha$ in the protein sequence), a compact unfolded one, and an elongated unfolded configuration.

To be able to quantify the degree of convergence of the free energy landscapes reconstructed by Monte Carlo metadynamics, we calculate the standard deviation between the reconstructed landscape after $T$ steps and the reference one
\begin{equation}
\sigma = \sqrt{\frac{1}{R_{max}-R_{min}} \int^{R_{max}}_{R_{min}} \left( f_{meta}(x) - f_{MC}(x) \right)^2 dx} \;,
\end{equation}
where $f_{meta}(x)$ and $f_{MC}(x)$ are the reconstructed and reference landscapes respectively, while $R_{min}$ and $R_{max}$ define the range of the RMSD over which $\sigma$ is calculated. They are chosen so as to englobe in the calculation the most significant fraction of the landscape, the corresponding values being 2 and 16 \AA, respectively. The reason why the edges of the landscape are not included in the calculation is that they are both noisy, as they are seldom visited by the system, aside from corresponding to high values of the free energy ($\gg k\theta$, where $k$ is the Boltzmann constant and $\theta$ is the temperature), and consequenlty not interesting from the thermodynamic point of view.

The two collective variables used are the RMSD and the radius of gyration $R_g$. We then proceed to integrate out the radius of gyration
\begin{equation}
F(RMSD)=-\theta\log\int dR_g\;\exp\left[\frac{-F(RMSD,R_g)}{\theta}\right] \;,
\end{equation}
where $\theta$ is the simulation temperature, so as to have a simpler, one--dimensional, visualization of the free energy landscape. The simulations are carried out at different values of the height $W$ of the Gaussian terms added to the non--Markovian potential and of their deposition time $\tau$, with each Gaussian having a fixed standard deviation $\delta s = 0.25$.

\begin{figure}[htb!]
\begin{center}
\includegraphics*[width=0.45\textwidth]{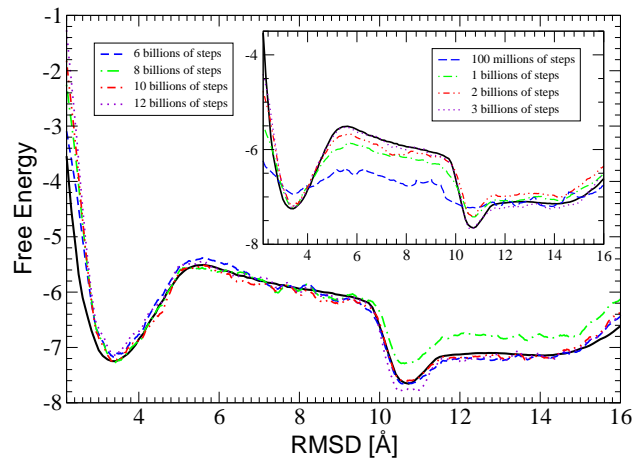}
\caption{Expected and reconstructed energy landscapes for a typical simulation. Shown in continuous line is the expected energy landscape. It can be appreciated how, after having reached the minimum value for $\sigma$, the reconstructed landscapes (here shown one every 2 billions of steps, starting from 6 billions) have small oscillations around the expected one. (Inset) Expected and reconstructed energy landscapes, ranging from the reconstructed landscape after about 100 millions of steps, to the one at the minimum $\sigma$, after 3 billions of steps.}
\label{fig:landscapes}
\end{center}
\end{figure}

In Fig. \ref{fig:landscapes} we show the reconstructed free--energy landscape as a function of the RMSD at different values of the number of MCS elapsed in the simulation (with $W = 0.01$ and $\tau = 200000$).
At the beginning (see the inset) the protein explores mainly the regions around 3 and 11 \AA, producing the two minima associated with the native and the denatured state. After these have been filled by the non--Markovian term, the rest of the landscape is refined and converges to the reference one within $3 \times 10^9$ MCS (to be compared with the $8 \times 10^{10}$ MCS needed by the standard Monte Carlo simulation).

\begin{figure}[htb!]
\begin{center}
\includegraphics*[width=0.45\textwidth]{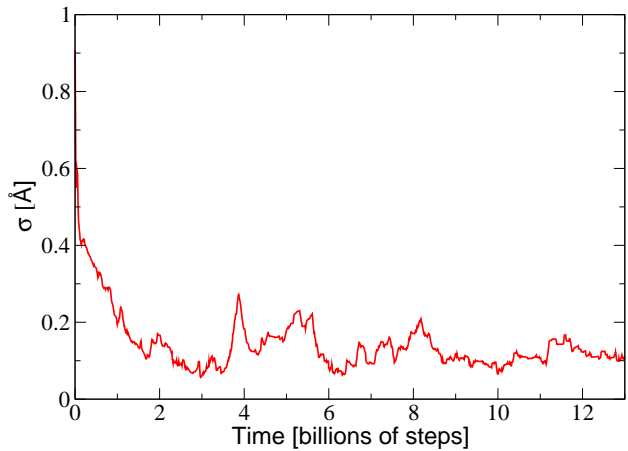}
\caption{$\sigma$ as a function of time. It can easily be seen how it goes down really fast, and then starts to oscillate in a quite small zone.}
\label{fig:sigma_vs_t}
\end{center}
\end{figure}

The corresponding values of $\sigma$ are displayed in Fig. \ref{fig:sigma_vs_t} as a function of the number of MCS.
The simulation reaches a fairly low $\sigma$ (under 0.2 \AA) in a few billions MCS, and then oscillates (with a spread of less than $\sim 0.15$ \AA) around $\sim 0.1$ \AA.

\begin{figure}[htb!]
\begin{center}
\includegraphics*[width=0.45\textwidth]{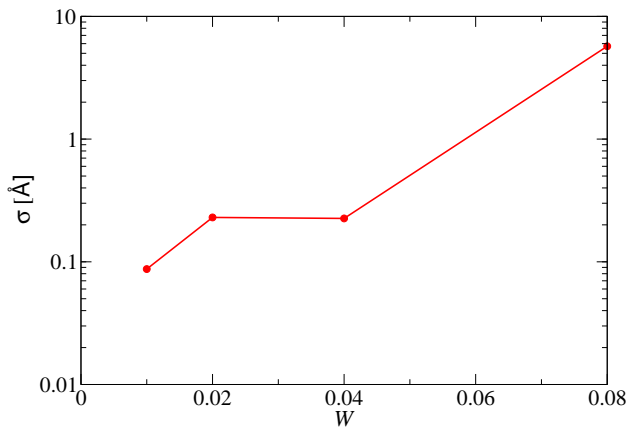}
\caption{$\sigma$ as a function of $W$ for a fixed $\tau$.}
\label{fig:sigma_vs_h}
\end{center}
\end{figure}

Fig. \ref{fig:sigma_vs_h} shows the dependence of $\sigma$ on $W$, at fixed $\tau = 200000$. The value of $\sigma$ plotted here corresponds to the average of the last $2 \times 10^6$ MCS of the simulation (cf. Fig. \ref{fig:sigma_vs_t}). The plot shows a steep increase of $\sigma$ with respect to the height of the hills (note the logarithmic axis scale), indicating that only a fine--grained deposition of the non--Markovian term is able to drive the system to equilibrium. This is consistent with the fact that Eq. (\ref{eq_du}) is derived by Eq. (\ref{eq_ust}) as an expansion for small $W$, and consequently fails when $W$ is increased.

\begin{figure}[htb!]
\begin{center}
\includegraphics*[width=0.45\textwidth]{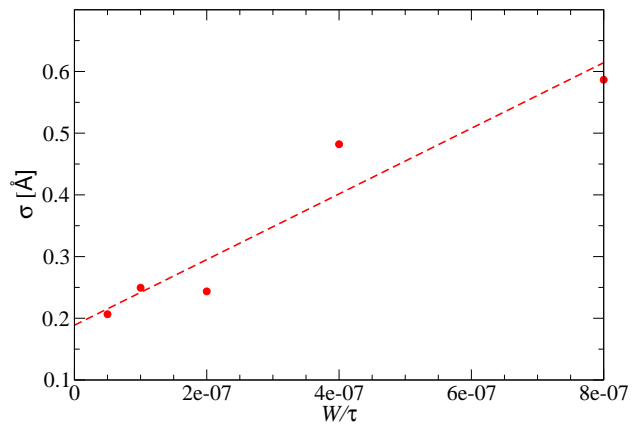}
\caption{$\sigma$ as a function of $W/\tau$. (dashed, red online) Linear fit, with intercept 0.189, slope 5.319e+05, standard deviation 0.170 and correlation coefficient 0.955.}
\label{fig:sigma_vs_h_on_dt}
\end{center}
\end{figure}

In Fig. \ref{fig:sigma_vs_h_on_dt} the dependence of $\sigma$ on $W/\tau$ is shown. It was shown by Laio and coworkers (\cite{Lai.ea:05}) that the lower is $W/\tau$ , the higher the accuracy of standard metadynamics is. The data shown indicate that also in Monte Carlo metadynamics the accuracy of the reconstructed landscape increases when $W/\tau$ is decreased. In particular, it seems that $\sigma$ is related to this ratio by a linear function (the linear fit indicates a slope of $5.3\times 10^5$, with a correlation of 0.955).

\begin{figure}[htb!]
\begin{center}
\includegraphics*[width=0.45\textwidth]{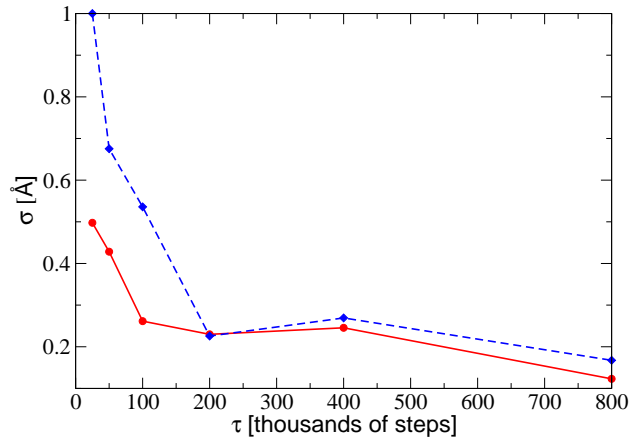}
\caption{$\sigma$ as a function of $\tau$ for two different, fixed, $W$. In particular, $W=0.02$, continuous line (red online) and $W=0.04$, dashed line (blue online).}
\label{fig:sigma_vs_dt}
\end{center}
\end{figure}

Fig. \ref{fig:sigma_vs_dt} shows how $\sigma$ varies for different values of $\tau$, at fixed $W$. As $\tau$ increases, $\sigma$ becomes smaller, as expected from the theorerical discussion carried out in Section \ref{theory}. In fact, the larger is $\tau$, the more likely it is for the system to having explored a region $A(T)$ exhaustively and thus for Eq. (\ref{eq_pst}) to be a good approximation of the actual probability distribution.

\section{Conclusion}

The metadynamics strategy has been implemented within a Monte Carlo scheme in order to take benifit from the positive aspects of both approaches. The algorithm is tested with a simplified protein model, and results particularly efficient and accurate in reconstructing the free energy landscape of the protein.

% \appendix

% \section{}
% \label{}

% Bibliographic references with the natbib package:
% Parenthetical: \citep{Bai92} produces (Bailyn 1992).
% Textual: \citet{Bai95} produces Bailyn et al. (1995).
% An affix and part of a reference:
%   \citep[e.g.][Ch. 2]{Bar76}
%   produces (e.g. Barnes et al. 1976, Ch. 2).

%\begin{thebibliography}{}

% \bibitem[Names(Year)]{label} or \bibitem[Names(Year)Long names]{label}.
% (\harvarditem{Name}{Year}{label} is also supported.)

% \bibitem[]{}

%\end{thebibliography}

\bibliographystyle{elsart-harv}
\bibliography{biblio}

\end{document}